\documentclass[a4paper]{article}
\usepackage{issp2020,amssymb,amsmath,graphicx}
\usepackage{filecontents}
\usepackage{graphicx}

\usepackage[utf8]{inputenc}
\usepackage[T1]{fontenc}
\usepackage[english]{babel} 
\usepackage[autostyle]{csquotes}
\usepackage{caption}
\usepackage{subcaption}
\usepackage[backend=biber,style=authoryear,sorting=nyt]{biblatex}
\defbibheading{refs}{\section{References}}
\ninept

\title{SPEAK WITH YOUR HANDS \\ Using Continuous Hand Gestures to control Articulatory Speech Synthesizer}

\makeatletter
\def\name#1{\gdef\@name{#1\\}}
\makeatother
\name{{ Pramit Saha$^1$, Debasish Ray Mohapatra$^1$, Sidney Fels$^1$}}

\address{\small \em $^1$HCT Lab, Department of Electrical and
Computer Engineering, University of British Columbia\\
{\small \tt pramit@ece.ubc.ca, debasishray@ece.ubc.ca, ssfels@ece.ubc.ca}}

\begin{document}
\maketitle

\begin{abstract}

This work presents our advancements in controlling an articulatory speech synthesis engine, \textit{viz.}, Pink Trombone, with hand gestures. Our interface translates continuous finger movements and wrist flexion into continuous speech using vocal tract area-function based articulatory speech synthesis. We use Cyberglove II with 18 sensors to capture the kinematic information of the wrist and the individual fingers, in order to control a virtual tongue. The coordinates and the bending values of the sensors are then utilized to fit a spline tongue model that smoothens out the noisy values and outliers. Considering the upper palate as fixed and the spline model as the dynamically moving lower surface (tongue) of the vocal tract, we compute 1D area functional values that are fed to the Pink Trombone, generating continuous speech sounds. Therefore, by learning to manipulate one's wrist and fingers, one can learn to produce speech sounds just through one's hands, without the need for using the vocal tract. 

\end{abstract}
\vspace*{1em}
\noindent \textbf{Keywords:} articulatory speech synthesizer, pink trombone, Cyberglove II, hand gestures, continuous vowel synthesis, kinematics-to-articulatory mapping, silent speech interface

\section{Introduction}
Articulatory speech synthesis \parencite{saha2018sound} encompasses the production of speech sounds using a vocal tract model and simulating the movements of the speech articulators like tongue, lips, velum etc. Articulatory vocal tract modelling targets the simulation of the process of speaking by recreating the behaviour of the human speech apparatus. 

Among the different articulators, tongue is the most dynamic as well as the most significant part of vocal tract. Hence, in this work, we attempt to control the tongue movements via wrist and finger movements, targeting vocal sound synthesis. This alters the anterior
part of the upper airway, thereby modulating sound propagation
through it. The goal is to develop a convenient, easy-to-learn and intuitive physical interface with improved control, leveraging the multi DOF capability of hand.

\section{Related previous works}

Recently, we have performed three interesting projects to synthesize speech from hand movements through articulatory and acoustic pathways. These three interfaces developed the background for the current project and hence are worth mentioning here.

The first one was about developing an interface named SOUND STREAM, involving five degree-of-freedom (DOF) mechanical control of a two dimensional, mid-sagittal human tongue-like structure (spring-based) for articulatory speech synthesis. As a demonstration of the project, the user was able to learn to produce a range of sounds, by varying the shape and position of the upper surface of the tongue-like structure in 2D space through a set of three sliders mounted on movable platform. The magnitude and frequency of the glottal excitation was controlled physically by two additional sliders. This entire arrangement allowed the user to play around with five sliders to vary the articulatory structures as well as the source acoustic parameters, exploring the variation of sounds.

The second version of the project was about developing another interface for articulatory speech synthesis named SOUND STREAM II \parencite{saha2018sound}- involving four DOF mechanical control of a two dimensional, mid-sagittal tongue through a biomechanical toolkit called ArtiSynth and a sound synthesis engine called JASS. As a demonstration of the project, the user learnt to produce a range of JASS vocal sounds, by varying the shape and position of the ArtiSynth tongue in 2D space through a set of four muscle excitors modeled using force-based sensors. This variation was computed in terms of Area Functions in ArtiSynth and communicated to the JASS audio-synthesizer coupled with two-mass glottal excitation model to complete this end-to-end gesture-to-sound mapping.

The goal of the third project was to develop a formant based vowel sound synthesizer \parencite{unknown} using CyberGlove as an input device to map continuous hand gestures (wrist flexion and extension; finger abduction and adduction) to English vowels. The interface enabled the user to control his wrist and finger movements (in a 1D + 1D control space) in order to synthesise a continuous vowel space (using first and second formants) easily and intuitively. This was motivated from the implementation of another adaptive speech interface named Glove Talk II \parencite{fels1998glove}, which achieved a neural network based mapping between continuous hand gestures and control parameters of a formant based speech synthesiser.


\section{Data collection}

\begin{figure}
	\centering
\includegraphics[scale=0.4]{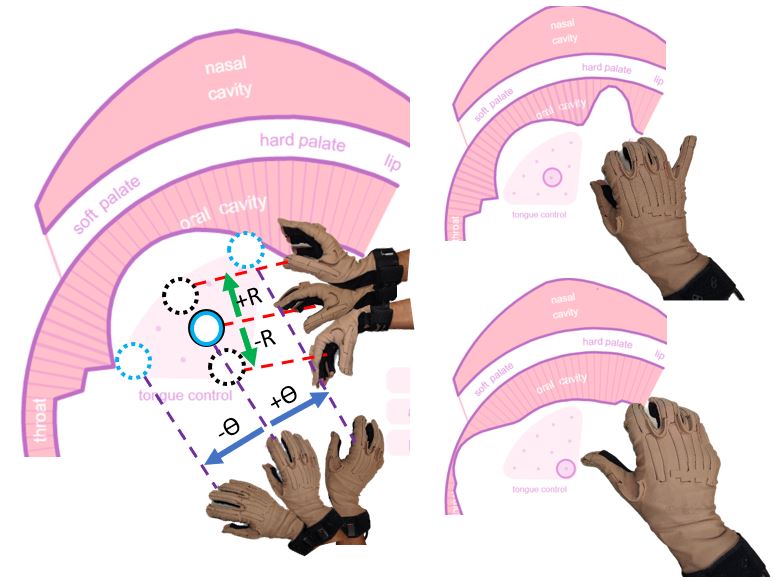}
\caption{The proposed pink trombone control paradigm}
\vspace{-10pt}
\end{figure}

We use CyberGlove II, manufactured by Immersion Inc., having 18 sensors (3 on each finger and 3 on the wrist), to capture hand movements. During the experiment, the arm of the participant is fixed at all instants. The participant is asked to perform wrist flexion and extension in order to decrease and increase the radius '$R$' as well as perform radial and ulnar deviation in order to change the angle '$\theta$' as shown in Fig 1. This is captured by the wrist sensors. Further the tip sensor of each of the 5 fingers is used to collect data on modulation of the shape of the tongue articulator. 

In the base position, the wrist is kept horizontal w.r.t the arm which is mapped to the mid-radius location in the vowel triangle in the Fig 1. Wrist flexion implies increase in jaw opening and tongue flattening where as, wrist extension implies decrease in jaw opening and the tongue elevation. The radial and ulner wrist deviations correspond respectively to retraction and protrusions of the tongue. The wrist is mainly used to make the front and back as well as high and low vowel sounds. 

Now, in order to accommodate few other changes in the tongue shape, we use the finger movements. In the base tongue shape, the fingers remain stationary in a gesture of gripping a ball as shown in Fig 1. Further, the fingers undergo extension accordingly, in order to change the tongue shape in a desired manner. For example, the elevation of the pinky finger to the correct degree can bring about alveolar fricatives like /s/, /z/, while particular elevation of the ring finger can produce palato-alveolar fricatives 
and correct elevation of the thumb can produce velar plosives like /k/, /g/ and so on. 
\section{Computation of area function}

The coordinates of the upper palate remain pre-specified and static. The sensor recordings from the wrist and the five fingers are used to fit a tongue spline model that changes dynamically with the changing tongue shape and position. The difference between the coordinates on the spline surface and the fixed palate is used as the diameter to compute the 1D oral tract area functional values as depicted in the literature \parencite{mathur2003vocal}. These area functions quantify the vocal tract shape.

\section{Articulatory speech synthesis engine}
The area functions are spatially sampled such that the vocal tract can be modelled as a series of adjoining cylindrical tube elements. The physics-based articulatory speech synthesizers model the acoustic wave propagation through these elements. We feed the area functional values to the Articulatory speech synthesis engine, named Pink Trombone \parencite{pt}, an online voice synthesis based web application that presents an interactive mid-sagittal view of human vocal tract and can be manipulated by users to simulate various vocal sounds. Out of numerous methods, the digital waveguide-based acoustic wave solvers can precisely compute acoustic wave propagation with an improved time performance. This interface maps the 1D Area functions to the acoustic space dynamically by forming a representation of such wave propagation. It also uses a glottal wave derived from LF-model to induce the acoustic energy. 

\section{Continuous vocal sound generation}
The user can constantly manipulate his wrist and finger movements to make the desired tongue shape and position. This dynamically changes the area function. The ability of Pink Trombone to continuously map the varying area functions allows the generation of a connected chain of speech sounds forming continuous utterances. This interface particularly enables the user to generate all the vowels, oral stop consonants, affricates, fricatives as well as approximants. Learning to control the fingers properly over time will enable the user to continuously synthesize meaningful words and sentences using these components.
\section{Discussion and Conclusion}
This work aims at developing a better physical user-interface through multi-degree of freedom control arrangement of human wrist and fingers. It can be easily extended to control the vocal fold parameters (like pitch), the lip movements and the nasalisation by using more DOF control strategies. This would lead to incorporation of more features into the interface and would make it more naturally sounding. 

However, one of the main fundamental questions that still remains is, how to quantify the usability, effectiveness and efficiency of the proposed interface. In other words, how to design a Fitts' task with varying level of difficulties which will help us find an equivalent index of difficulty or the information rate in bits/second required by the proposed hand gesture control paradigm. 
We have observed that making fricatives like /s/, /z/ is a much more difficult task (where it needs better precision regarding placement of the tongue tip - at a particular distance vertically downwards from the hard palate) than making stop consonants like /t/ and /d/ (where the tongue tip or body has to  strike anywhere within a wider range of positions directly on the hard palate). Taking into account these considerations, we will continue to investigate how difficult it is for the wrist and finger trajectories to make similar sounds and thereby will get better insights on designing proper experiments for user-study. 

\section{Acknowledgements}
This work was funded by the Natural Sciences and Engineering Research Council (NSERC) of Canada.

\eightpt
\printbibliography[heading=refs]
\end{document}